\documentclass[aps,pra,twocolumn,showpacs,prabib]{revtex4}
\usepackage{graphicx}
\usepackage{amsmath}

\newcommand{\bea}{\begin{eqnarray}}
\newcommand{\eea}{\end{eqnarray}}
\newcommand{\be}{\begin{equation}}
\newcommand{\ee}{\end{equation}}

\begin{document}

\title{Direct Relation between the Excess Current Noise Spectral Density and
the Complex Conductivity of Disordered Conductor}

\author{Peter V. Pikhitsa}
\affiliation{Seoul National University, 151-742 Seoul, Korea}

\begin{abstract}

We put forward a fundamental relationship between the normalized spectral
density of excess current noise and the imaginary part of
complex conductivity.
In the case of metal our expression enables us to describe reliable experimental data on the excess noise giving nearly $1/f$ spectrum and correct temperature dependence of the Hooge factor.
In the case of non-Debye conductors such as semiconductors and ionic glasses it leads to $1/f^s$ ($s<1$) for the excess noise spectrum. It is demonstrated that reactive "capacitance" effects in electron conductivity are responsible for the excess current noise.
The general relation that we introduce turns out to be a direct consequence of the continuous time random walk model.

\end{abstract}


\maketitle

Many attempts have been made to explain the origin of excess current noise
with the low-frequency spectrum $1/f$ on the background of direct current in
various conducting materials \cite{1,2,3}. It is believed to have been
established that $1/f$ noise is determined by some relaxation processes
accompanying the movement of carriers in disordered conductor. For example,
in metals electrons are scattered by relaxing defect structures \cite{4,5}
while in amorphous semiconductors and ionic glasses relaxation can be
connected with trapping and hopping mechanisms of conductivity \cite{6,7}.

A broad distribution of relaxation times which usually have an activation
form of dependence upon local energy barrier heights is very important for
obtaining $1/f$ spectrum \cite{1,2,3}. On the other hand, such a
distribution inevitably leads to the low-frequency dependence $\omega ^s$ (where $0<s<1$ and $\omega=2\pi f$) of ac conductivity \cite{7} which is well-known for amorphous
semiconductors and solid electrolytes \cite{8,9,10}. Therefore, bearing
in mind the common origin for both phenomena, one should expect the
existence of some relation that links together the spectral density of the
excess current noise and the frequency dependent complex conductivity in
disordered conductors.

As it was stressed \cite{10a} for flicker noise there must exist "...
intimate connection with conductivity ... a close relationship between the
forms of the flicker noise and unsimulated noise...". Previous attempts to
establish it were called "untenable" \cite{2}.

In the present paper we find this relation with the help of the
well-known continuous time random walk (CTRW) model \cite{11,12,13}. Considering direct consequences of the revealed relation we demonstrate its applicability for the description of reliable experimental data. This allows one to make a conclusion that in spite of the model way of its derivation this relation is of fundamental nature.

Indeed, the CTRW model gives for the normalized excess current noise spectral density $%
S(\omega )$ the expression
\begin{eqnarray}
S(\omega )=\frac{NS_I(\omega )}{I^2}=4\tilde t\mbox{Re}\left[ \frac{\tilde
\psi (i\omega )}{1-\tilde \psi (i\omega )}\right] ,
\end{eqnarray}
where $S_I(\omega )$ is the noise spectral density, $I$ is the average
current, $N$ is the number of carriers, $\tilde \psi (i\omega )$ is the
Fourier transform \cite{6} of the waiting-time distribution function $\psi
(t)$ and $\tilde t=\int_0^\infty t\psi (t)dt$. The complex conductivity
obtained in the CTRW model has the form \cite{6}
\begin{eqnarray}
\hat \sigma (\omega )=\sigma (0)\tilde t\frac{i\omega \tilde \psi (i\omega )
}{1-\tilde \psi (i\omega )},
\end{eqnarray}
where $\sigma (0)$ is the conductivity at zero frequency (the so-called dc
conductivity). From Eqs.(1) and (2) it immediately follows that
\begin{eqnarray}
S(\omega )=\frac 4\omega \frac{\mbox{Im}\hat \sigma (\omega )}{\sigma (0)}.
\end{eqnarray}

One may argue that the CTRW model we have used is oversimplified. However, those consequences from Eq.(3) that we obtain in the present work enables us to conclude that it is  not restricted by the
limitations of the CTRW model which Eq.(3) has been derived from.

To show this let us apply Eq.(3) to the case of metal with small
concentration $c_D=n_D/n_a\ll 1$ of randomly distributed point defects,
where $n_D$ is the number density of the defects and $n_a$ is the number
density of atoms. Let each single defect have $g$ states (for example,
orientational ones \cite{4,5}) which are characterized by different
transport cross-sections $\sigma _i$ ($i=1,2\dots g$) of electron scattering
on a defect and let the number density of defects in $i$th state be $n_i$,
so that
\begin{eqnarray}
\sum_{i=1}^gn_i=n_D.
\end{eqnarray}
Following \cite{5} we assume that transitions between the states of a defect
occur through an energy barrier $E_0+\delta E$ which contains the random
component $\delta E$ caused by non-uniform strain fields quenched in the
polycrystal. We take for $\delta E$ the Gaussian distribution function
\begin{eqnarray}
P(E)=\frac 1{\sqrt{2\pi T_{*}}}\exp \left[ -\left( \frac{\delta E}{2T_{*}}%
\right) ^2\right] ,
\end{eqnarray}
where $T_{*}$ is some characteristic temperature for the quenched disorder
\cite{5}. Together with the activation form of relaxation times $\tau
_r=\tau _0\exp [(E_0+\delta E)/T]$ Eq.(5) leads to the log-normal
distribution $P(\tau _r)$ of these times.

Let us write down the real part of the complex conductivity for a high
frequency (greater than inverse relaxation times but less than phonon
frequencies). For this case all defect states can be considered randomly
''quenched'' as far as electrons do not "see" the changes in defect states.
Then following \cite{14}
\begin{multline}
\mbox{Re\,}\hat \sigma (\infty )=\\
\left\langle \left[ \rho _{ph,e}(T)+\frac{%
p_F }{e^2}<\sigma >\frac{n_D}{n_e}+\frac{p_F}{e^2}\sum_{i=1}^g\sigma _i\frac{%
\Delta n_i}{n_e}\right] ^{-1}\right\rangle ,
\end{multline}
where beside the temperature-dependent contribution to resistivity $\rho
_{ph,e}(T)$ from phonon- and electron-electron scattering there appear terms
of residual resistivity \cite{14} and we introduced $\Delta n_i=n_i-n_D/g$; $%
p_F$ is the Fermi momentum, $e$ is the electron charge, $n_e=n_a$ is the
number density of electrons, $<\sigma >=\left( \sum_{i=1}^g\sigma _i\right)
/g$. For metals $c_D\sim 10^{-4}$ \cite{15} is a small expansion parameter
so that having expanded Eq.(6) up to quadratic terms with respect to
residual resistivity fluctuations we find
\begin{eqnarray}
\mbox{Re\,}\hat \sigma (\infty )=\frac 1{\rho (0)}+\frac{p_F^2}{e^4\rho (0)^3%
}(\Delta \sigma )^2c_D\left( 1-\frac 1g\right) ,
\end{eqnarray}
where we denoted $\rho (0)=\rho _{ph,e}(T)+p_F<\sigma >c_D/e^2$ - the
total resistivity at $\omega =0$, and
\[
(\Delta \sigma )^2=\frac 1g\sum_{i=1}^g\sigma _i^2-\frac
1{g(g-1)}\sum_{i\neq j=1}^g\sigma _i\sigma _j,
\]
and we used the result of averaging over the random spatial distribution of
defects \cite{5}: $<(\Delta n_i/n_e)^2>=c_D(1-1/g)/g$.

Let us make it clear the origin of the difference between high and low frequency dependence of conductivity due to defects. For example, if one has a resistor network then at zero frequency the conductivity is determined by the average resistivity proportional to
the successive sum of local resistivity in each part of the network. Just the quantity which is reciprocal to the average specific
resistivity can be identified with the experimentally observable
conductivity of such a system that is $\sigma(0)=1/<\rho>$ and brackets denote the abovesaid averaging. Meanwhile, the high frequency conductivity is a "parallel channel" one in nature so that $\sigma(\infty)=<1/\rho>$ like in Eq.(6). These two quantities coincide only in the absence of spatial fluctuations of resistivity in the resistor network. These fluctuations are analogous of a defect distribution in the case of disordered metal.

Thus for our case at high frequency the increase in average conductivity is due to regions with low defect concentration (voids of low resistivity) unlike the dc case when the high and low resistivity regions sum up successively. It implies that reactive "capacitance" effects in electron conductivity are the origin of the dispersion of the conductivity leading to the excess noise according to Eq.(3).

In the case of a distribution of relaxation times the defects with $\tau_r>1/\omega$ can be considered completely
"quenched" and the defects with $\tau_r<1/\omega$ - completely relaxed. Let
us introduce the fraction $X(\omega)$ of defects which can be "seen" by
conduction electrons as "quenched" in their random states:
\[
X(\omega)=\int_{1/\omega}^\infty P(\tau_r) d \tau_r.
\]
For high frequency $X(\omega)=1$ and for low frequency it turns to zero.

Then we can generalize Eq.(7) for an arbitrary frequency $\omega $
\begin{eqnarray}
\mbox{Re\,}\hat \sigma (\omega )=\frac 1{\rho (0)}+\frac{p_F^2}{e^4\rho (0)^3%
}(\Delta \sigma )^2c_D\left( 1-\frac 1g\right) X(\omega ),
\end{eqnarray}
so that for $\omega \rightarrow \infty $ Eq.(8) coincides with Eq.(7) and
for $\omega \rightarrow 0$ it gives $\mbox{Re}\widehat{\sigma }(0)=1/\rho (0)
$ as it should be.

For the log-normal distribution function $P(\tau_r)$ one can obtain
\begin{eqnarray}
X(\omega)=\frac12\left[1+\mbox{erf} \left(\frac{w}{\sqrt{\nu}}\right)\right],
\end{eqnarray}
where erf is the error function, $w=\ln \omega\tau_{r0}$, $\tau_{r0}=\tau_0
\exp(E_0/T)$ and $\nu=4T_*^2/T^2$ is the fluctuation exponent introduced in
\cite{7}. It is important for what follows that for $T_*\sim 1000$ K \cite{5}
and $T<600$ K \cite{1} $\nu\gg 1$.

Let us introduce a function
\begin{eqnarray}
\hat z_1(\omega )=\frac 1{\sqrt{\pi \nu }}\int_{-\infty }^\infty \frac{\exp
\left( -{u^2}/{\nu }\right) }{i\omega \tau _{r0}+\exp (u)}du
\end{eqnarray}
investigated in \cite{7} and for which in the limit $\nu \gg 1$ there holds
the formula
\begin{eqnarray}
\mbox{Im\,}\hat z_1(\omega )=-\frac{e^{-w}}2\left[ 1+\mbox{erf}\left( \frac
w{\sqrt{\nu }}\right) \right] .
\end{eqnarray}
Taking into account Eqs.(9) and (11) one can rewrite Eq.(8) as
\begin{multline}
\mbox{Re\,}\hat \sigma (\omega )=\\
\frac 1{\rho (0)}-\frac{p_F^2}{e^4\rho (0)^3%
}(\Delta \sigma )^2c_D\left( 1-\frac 1g\right) \omega \tau _{r0}\mbox{Im\,}%
\hat z_1(\omega )
\end{multline}
and after using the Kramers-Kronig relation \cite{16} we have
\begin{multline}
\mbox{Im\,}{\hat \sigma (\omega )}=\\
\omega \tau _{r0}\frac{p_F^2}{e^4\rho
(0)^3}(\Delta \sigma )^2c_D\left( 1-\frac 1g\right) \mbox{Re\,}\hat
z_1(\omega ).
\end{multline}
It is worth noting that although the frequency dependent term in Eq.(12) is
negligibly small comparatively to $\sigma (0)=1/\rho (0)$ it is this term
that gives all the effect in Eq.(13). Substituting Eq.(13) in Eq.(3) we
finally have
\begin{multline}
S(\omega )=
\frac{p_F^2}{e^4\rho (0)^2}(\Delta \sigma )^2c_D\left( 1-\frac
1g\right)\\
 \mbox{Re}\left[ \frac{\tau _{r0}}{\sqrt{\pi }}\int_{-\infty
}^\infty \frac{\exp \left( -{u^2}\right) }{i\omega \tau _{r0}+\exp
(2T_{*}u/T)}du\right] .
\end{multline}

This equation exactly coincides with the corresponding Eq.(27) from \cite{5}
(where Eq.(27) was derived by direct calculation of the temporal correlation
function of resistivity) if one calculates $(\Delta \sigma )^2$ taking into
account the orientation modes of defects assumed in \cite{5}. The important
difference between Eq.(14) and Eq.(27) from \cite{5} is that Eq.(14) is more
general as to the assumptions made about the specific structure of defects.
Namely, the Eq.(27) from \cite{5}, in fact, describes the spectral density of anisotropic
current fluctuations only.
As far as Eq.(14) was studied in detail in \cite{5} and was shown to give $%
1/f$ noise spectrum and a very accurate temperature dependence of the Hooge factor for
metal we do not repeat these results here.

The other interesting and non-trivial case where Eq.(3) can be used is the
case of strongly disordered conductors with a strong frequency dependence of
$\mbox{Re} \hat \sigma(\omega)$, namely, amorphous semiconductors and ionic glasses.
Here one should not expect a superposition form of Eq.(14) but has to use a
general approach (e.g. the one developed in \cite{7} for conductivity of
ionic glasses) to calculate $\hat \sigma(\omega)$ in matters with hopping
conductivity.

The conductivity $\hat \sigma(\omega)$ standing in Eq.(3) can be determined through the
complex specific impedance $\hat z(\omega)$ calculated in \cite{7} by the
relation
\begin{eqnarray}
\hat \sigma(\omega)=\frac{1}{\hat z^*(\omega)}-i\frac{\varepsilon_\infty
\omega}{4\pi}= \frac{\varepsilon_\infty}{4\pi}\left[\frac{1}{\tau_{r0}\hat
z_1(\omega)} -i\omega\right],
\end{eqnarray}
where $\varepsilon_\infty$ is the high-frequency dielectric constant of the
conductor and the complex conjugation was used to alter the sign before $%
i\omega$ so that the definitions of Fourier transforms in Eq.(3) and in \cite
{7} are put in accordance. In Figure 1 we give for illustration the result of
computation of $S(\omega )$ obtained according to Eqs.(3) and (15) with $%
\hat z(\omega )$ taken from Eq.(37) in \cite{7}. The parameters $\nu =10$
(so that corresponding $s=0.47$ according to Eq.(75) of \cite{7}), $%
\varepsilon _\infty =7$, and $\sigma (0)=10^{-8}$ $\Omega ^{-1}$ cm$^{-1}$
chosen for the calculation are close to ordinarily used for ionic glasses.
One can see from log-log scale Fig.1 that there exists the excess noise
spectrum: $1/\omega ^s$ (certainly, at very high frequencies it gives $%
1/\omega ^2$, not shown). For the parameters chosen here only the branch of $%
1/\omega ^s$ type can be observable in experiment. It is interesting to
notice that $1/\omega ^s$ spectrum with $s<1$ was predicted in \cite{17,18}
where it was discussed in relation with the non-Debye relaxation phenomena
(see also \cite{19,20} where $s$ was found to be $1/2$). For semiconductors
the value of $s$ for ac conductivity may be very close to unity.

It is useful to note that the tendency of the spectral density towards "saturation" in the
low-frequency limit (see Fig. 1) produces convergence of the noise integrated
intensity, so the "paradox" of the divergence of the mean square of the fluctuations of a
quantity responsible for the $1/f$ noise frequently discussed in the literature (see, for example,
the review \cite{3}) does not arise at all.

In conclusion, we have shown that there exists a relation given by Eq.(3)
between the excess noise spectrum and the complex conductivity. This
relation links together two phenomena previously considered separately, namely, the
 excess current noise and the frequency response in
disordered conductors. In spite of the fact that this relation has been
established with the help of the CTRW model we claim that it is fundamental
and there are some weighty reasons to find its proof on the basis of general
physical principles. We believe that our relation breaks "... a long tradition of models being proposed that are forgotten after a short time." (see \cite{17} where these words of F.N. Hooge were cited) .

   \acknowledgments

I would like to thank Prof. M.B. Weissman for helpful discussions.

\begin{figure}[h]
\centerline{\includegraphics[width=0.9\linewidth,clip=]{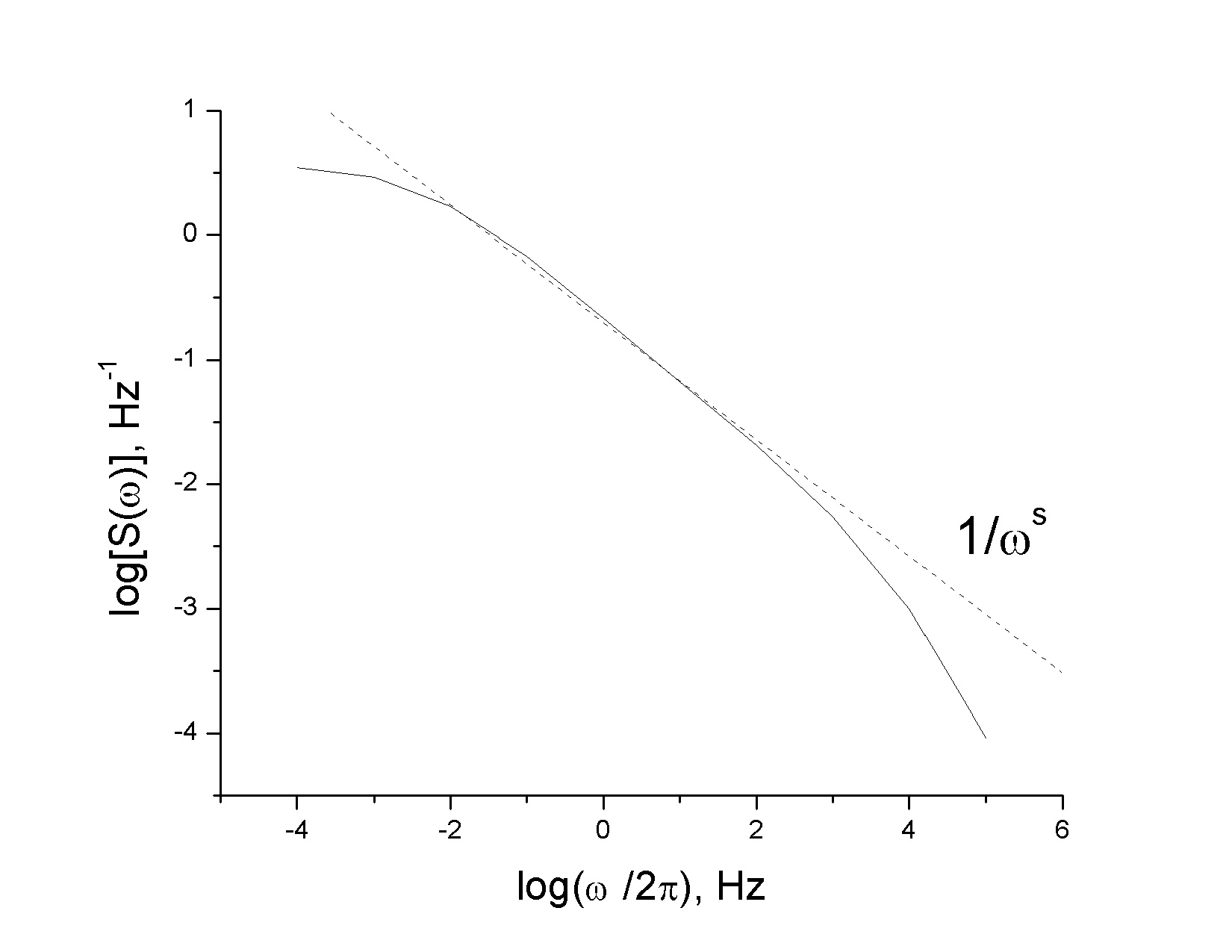}}
\caption{The noise spectrum of a disordered ionic conductor.
Parameters are given in text.}
\label{fig:fig1}
\end{figure}

\end{document}